\documentclass{PoS}

\usepackage{placeins}
\usepackage{tabularx}
\usepackage{amsmath}

\newcommand{\rmii}[1]{{\mbox{\tiny\rm{#1}}}}
\newcommand{\beq}{\begin{equation}}
\newcommand{\eeq}{\end{equation}}
\newcommand{\bea}{\begin{align}}
\newcommand{\eea}{\end{align}}
\newcommand{\beas}{\begin{align*}}
\newcommand{\eeas}{\end{align*}}

\newcommand{\eps}{{\varepsilon}}
\newcommand{\mD}{m_\rmii{D}}

\newcommand{\vE}{\vec E}
\newcommand{\rh}{\hat r}
\newcommand{\vr}{\vec r}
\newcommand{\vp}{\vec p}

\newcommand{\Tint}[1]{{\hbox{$\sum$}\!\!\!\!\!\!\!\int\,}_{\!\!\!\!\raise-0.9ex\hbox{$\scriptstyle{#1}$}}}

\title{A gauge invariant Debye mass for the complex heavy-quark potential}

\ShortTitle{A gauge invariant Debye mass for the complex heavy-quark potential}

\author{Yannis Burnier \\
        Institut de Th\'eorie des Ph\'enom\`enes Physiques, Ecole Polytechnique F\'ed\'erale de Lausanne, CH-1015, Lausanne, Switzerland,\\
        E-mail: \email{yannis.burnier@epfl.ch}}
  
\author{\speaker{Alexander Rothkopf}\\
        Institute for Theoretical Physics,  Heidelberg
  University, Philosophenweg 16, D-69120 Heidelberg, Germany\\
        E-mail: \email{rothkopf@thphys.uni-heidelberg.de}}      

\abstract{The concept of a screening mass is a powerful tool to simplify the intricate physics of in-medium test charges surrounded by light charge carriers. While it has been successfully used to describe electromagnetic properties, its definition and computation in QCD is plagued by questions of gauge invariance and the presence of non-perturbative contributions from the magnetic sector. Here we present a recent alternative definition of a gauge invariant Debye mass parameter following closely the original idea of Debye and H\"uckel. Our test charges are a static heavy quark-antiquark pair whose complex potential and its in-medium modification can be extracted using lattice QCD. By combining in a generalized Gauss-Law the non-perturbative aspects of quark binding with a perturbative ansatz for the medium effects, we succeed to describe the lattice values of the potential with a single temperature dependent parameter, in turn identified with a Debye mass. We find that its behavior, as evaluated in a recent quenched lattice QCD study, deviates from that in other approaches, such as hard-thermal-loop perturbation theory or from electric field correlators on the lattice. In particular around the phase transition its values tend to zero significantly faster than at weak-coupling.}

\FullConference{34th annual International Symposium on Lattice Field Theory\\
		24-30 July 2016\\
		University of Southampton, UK}

\begin{document}

The Debye mass in Quantum-Electrodynamics (QED) is a well defined concept and related to the decay of the gauge invariant static electric field correlator in a thermal medium or equivalently to the dynamical emergence of a mass for longitudinal photons from their static self energy $\Pi$
\begin{align}
\langle {\bf E}({\bf x}) {\bf E}({\bf 0}) \rangle_{\rm T>0} \sim e^{-m_D^{\rm QED} | {\bf x}|}/|{\bf x}^3|, \quad \lim_{{\bf p}\to0} \Pi_{00}=\big(m_D^{\rm QED}\big)^2, \quad \lim_{{\bf p}\to0} \Pi_{ij}=0
\end{align}
Unfortunately neither the color electric field nor the gluon self energy are gauge invariant in non-Abelian gauge theories. Perturbatively it is only possible to compute the self energy in Quantum-Chromodynamics (QCD) \cite{Arnold:1995bh} up to logarithmic correction to the leading order $m_D^{\rm LO}=gT\sqrt{\frac{N_c}{3}+\frac{N_f}{6}}$, 
\begin{align}
m_D^{\rm QCD}=m_D^{\rm LO} + \frac{N_c g^2 T}{4\pi}{\rm log}[\frac{m_D^{\rm LO}}{g^2T}] + \kappa_1 Tg^2+\kappa_2 Tg^3\label{Eq:mDpert}
\end{align}
beyond which genuine non-perturbative effects from the magnetic sector contribute. Conventionally these are encoded as two additional constants $\kappa_{1,2}$, which need to be fixed using numerical simulations at high temperatures, where higher corrections in $g$ can be neglected. Formally the Debye mass may be defined from the gauge invariant correlator of color singlet operators $\epsilon_{ijk}{\rm Tr}[A_0F_{jk}]$, which are amenable to simulation in a 3-dimensional effective field theory called EQCD \cite{Kajantie:1997pd,Hart:2000ha,Laine:2009dh}. In this theory, which, if its matching to QCD is carried out perturbatively, is applicable at high temperatures $T>T_C$, rather sizable corrections $\kappa_1=2.5(2)$ and $\kappa_2=-0.5(2)$ have been found.

On the other hand one can consider the practical approach of returning to the original ideas of Debye and H\"uckel and to define the Debye mass by a physical process, e.g.\ the interaction of heavy test charges in a medium. While in the original work \cite{DebyHueckel} ions in solutions were considered, for QCD, heavy-quark bound states may play a similar role. Indeed it is known that the behavior of so called quarkonium ($c\bar{c}$ or $b\bar{b}$) can be described by an in-medium potential. Therefore if one could formulate the temperature dependence of such a potential in terms of a Debye mass it would provide a practical means to describe the relevant physics of in-medium quarkonium binding. Earlier works using model potentials \cite{Maezawa:2007fc,Digal:2005ht}, such as the color singlet free energies and fitting a purely screened Coulombic ansatz $-\alpha_S {\rm exp}[-m_D r]/r$ to their large distance behavior found quite small corrections $\kappa_1\approx0.35$ and $\kappa_2\approx-0.1$ to the perturbative behavior of $m_D^{\rm QCD}$.

\begin{figure*}[t]
\centering\vspace{-0.3cm}
\includegraphics[scale=0.35, trim= 0 2cm 0 1.9cm, clip=true ]{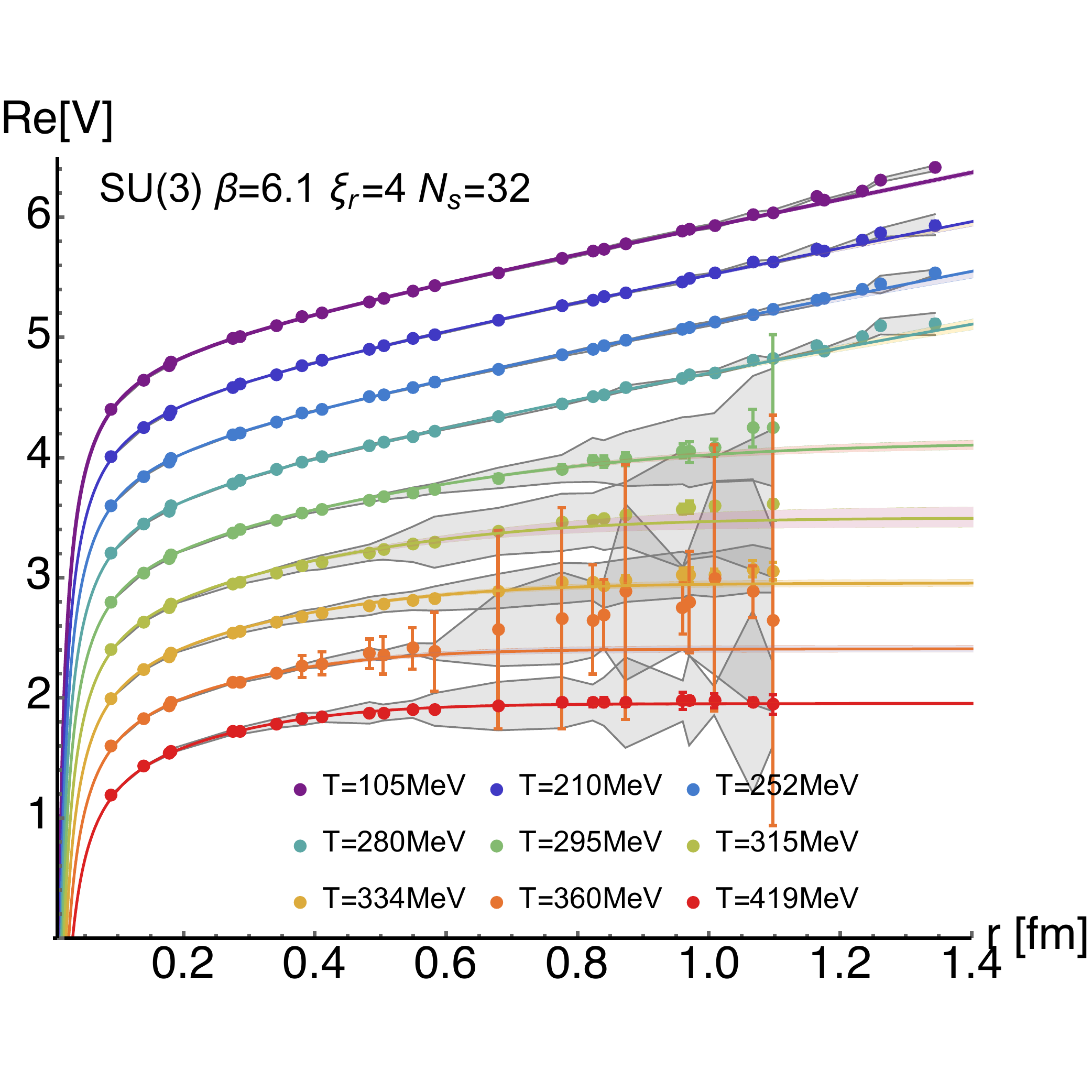}\hspace{0.2cm}
\includegraphics[scale=0.35, trim= 0 2cm 0 1.9cm, clip=true]{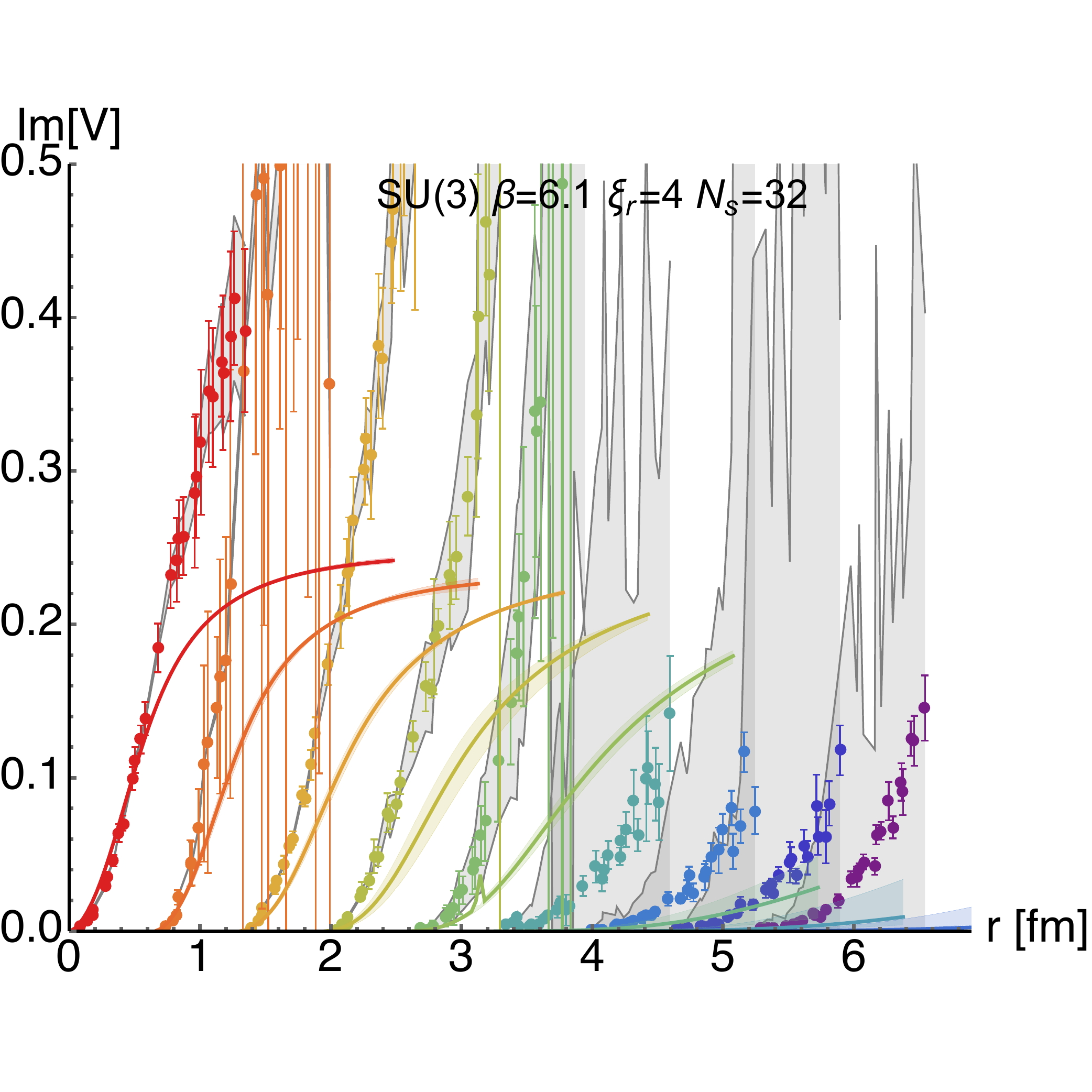}
\caption{${\rm Re}[V]$ (right) and ${\rm Im}[V]$ (left) extracted from quenched QCD simulations at $\beta=6.1$, $\xi_r=4$ on lattices with $N_s=32$ (filled points). The values of ${\rm Re}[V]$ are shifted by hand in y-direction, those of ${\rm Im}[V]$ in x-direction. Colored errorbars arise from a ten-bin Jackknife resampling and the gray errorbands from varying the default model of the underlying Bayesian spectral reconstruction. Solid lines show the analytic parametrization of the potential from a generalized Gauss-Law. Fitting ${\rm Re}[V]$ at $T=105$MeV establishes the vacuum parameters $\alpha_S$, $\sigma$ and $c$, while fits to ${\rm Re }[V]$ at $T\gtrsim T_C$ fix $m_D$. This $m_D$ then allows postdicting  {\rm Im}[V].}\label{SU3Pot}
\end{figure*}

Here we wish to go beyond these studies in two ways: first by using the proper effective field theory based potential extracted from lattice QCD \cite{Burnier:2016mxc} and secondly by elucidating its in-medium modification beyond simple coulomb screening \cite{Burnier:2015nsa}. The former is related to the fact that with the maturation of effective field theories, such a non-relativistic QCD (NRQCD) and potential NRQCD (pNRQCD) \cite{Brambilla:1999xf,Pineda:2000sz,Brambilla:2011sg} it has become possible to systematically relate the concept of an in-medium potential for heavy quarks at finite temperature with the underlying microscopic theory of QCD. Systematically here means that the potential can be written as a static potential with corrections according to higher powers of the heavy quark velocity $v\sim p/m$, which order by order can be matched to a QCD observable. If a potential description exists then the late time behavior of the Wilson loop can be used to define the static contribution via
\beq
V(r)=\lim_{t\to\infty} \frac{i\partial_t W_\square(t,r)}{W_\square(t,r)}, \quad W_\square(t,r)=\Big\langle {\rm Tr} \Big( {\rm exp}\Big[-ig\int_\square dx^\mu A_\mu^aT^a\Big] \Big) \Big\rangle\label{Eq:VRealTimeDef},
\eeq
which was evaluated for the first time in a resummed hard-thermal loop (HTL) perturbation theory in \cite{Laine:2006ns,Beraudo:2007ky} and shown to take on complex values
\beq
V^{\rm LO}_{\rm HTL}(r)= -\tilde\alpha_s\left[\mD+\frac{e^{-\mD r}}{r}
+iT\phi(\mD r)\right],\quad  \phi(x)=2 \int_0^\infty dz \frac{z}{(z^2+1)^2}\left(1-\frac{\sin(xz)}{xz}\right) \label{Eq:VHTL}.
\eeq
We absorb the factor $C_F$ into the coupling constant $\tilde\alpha_s=\frac{g_s^2C_F}{4\pi}$, i.e. use the same convention as in the phenomenology literature. Since a real-time definition such as Eq.\eqref{Eq:VRealTimeDef} is not directly amenable to lattice simulations it was proposed in \cite{Rothkopf:2009db,Rothkopf:2011db} to instead use a spectral decomposition of the Wilson loop to relate its spectral function to the potential
\begin{align}
V(r)=\lim_{t\to\infty}\int d\omega\, \omega e^{-i\omega t} \rho_\square(\omega,r)/\int d\omega\, e^{-i\omega t} \rho_\square(\omega,r). \label{Eq:PotSpec}
\end{align}
As shown more generally in \cite{Burnier:2012az} the potential description is applicable if a well defined lowest lying peak exists in the spectrum with its position and width being related to ${\rm Re}[V]$ and ${\rm Im}[V]$ respectively. The extraction of these spectra from lattice QCD simulations however required the development of a novel Bayesian approach \cite{Burnier:2013nla} before a first quantitatively robust result could be obtained \cite{Burnier:2014ssa}. In the following we will use an updated computation of the potential in quenched QCD \cite{Burnier:2016mxc} with naive anisotropic Wilson action on $32^3\times N_\tau$ lattices using $\beta=6.1$ and $\xi_r=4$ \cite{Matsufuru:2001cp} corresponding to an $a_s=0.097$fm and physical box size $L=3.1$fm. Using the fixed scale approach we span temperatures $T\in[105\ldots 419]{\rm MeV}=[0.39\ldots1.4]T_C$ by varying the number of temporal lattice sites $N_\tau\in[72\ldots20]$. The resulting ${\rm Re}[V]$ and ${\rm Im}[V]$ are shown on the left and right in Fig.\ref{SU3Pot} respectively. The values have been shifted by hand for better readability.

Note that below $T_C=290$MeV virtually no change in ${\rm Re}[V]$ can be seen, while just above $T_C$ its values show a qualitatively different behavior, with the linear rise being absent, asymptoting to a constant. This is reminiscent of the $SU(3)$ phase transition one would expect in the thermodynamic limit. The imaginary part of the potential also only shows significant values away from zero above $T_C$. 

Now our aim is to relate the in-medium modification of both ${\rm Re}[V]$ and ${\rm Im}[V]$ to the concept of a Debye mass \cite{Burnier:2015nsa}. The starting point is the observation that the real-valued low temperature potential between $r=0.1\ldots 1.2$fm can be extremely well described by a naive Cornell-type ansatz $V_{\rm T\approx0}(r)=-\alpha_S/r+\sigma r+c$, as seen from the topmost purple line on the left of Fig.\ref{SU3Pot}. The next step is to consistently describe the in-medium modification of this non-perturbative vacuum potential, which we attempt via the generalized gauss law \cite{Dixit:1989vq} for the auxiliary vector field $\vE$ we may associate with the static heavy quark potential
\beq
\vec\nabla \left(\frac{\vE}{r^{a+1}}\right)=   4\pi \,q\, \delta(\vr) \label{Eq:GenGauss}.
\eeq
This expression applies to a general $\vE=q r^{a-1} \rh$, which reduces either to the well known Coulombic $a=-1, q=\tilde\alpha_s, [\tilde\alpha_s]=1$ or the confining potential $a=1, q=\sigma, [\sigma]={\rm GeV}^2$. Originally Debye introduced a real-valued background charge to describe the in-medium behavior, which was used in \cite{Digal:2005ht} but which does not allow to treat ${\rm Im}[V]$. Instead we will introduce the medium via a permittivity, as is done e.g.\ in classical electrodynamics. We will make the ansatz of a non-perturbative bound state being immersed in a gas of weakly interacting quarks and gluons and hence use the permittivity from leading order HTL
\beq
\eps^{-1}(\vp,m_D)=\frac{p^2}{p^2+m_D^2}-i\pi T \frac{p\, m_D^2}{(p^2+m_D^2)^2}.\label{Eq:HTLperm}
\eeq
Note that the temperature enters solely via the single parameter $m_D$ here. This permittivity has been used to investigate the in-medium modification of the potential before \cite{Thakur:2013nia}, but in a way that lead to unphysical results (divergent ${\rm Im}[V]$ and unscreened component in ${\rm Re}[V]$). Our self consistent treatment using the Gauss-law avoids these issues.

For the Coulombic part of the vacuum potential we can straight forwardly transform Eq.\eqref{Eq:GenGauss} to momentum space and multiply $\epsilon^{-1}$ to the l.h.s.. Transforming back we obtain the linear-response like defining equation for the in-medium Coulomb potential
\beq
-\nabla^2 V_C(r)+m_D^2V_C(r)=\tilde\alpha_s \Big(  4 \pi  \delta(\vr)-  iT m_D^2 \varphi(m_Dr)\Big), \quad \varphi(x)=2\int_0^\infty dp  \frac{\sin(p x)}{p x}\frac{p}{p^2+1}\label{Eq:Coulomb(r)withIm},
\eeq
where the strength of the modification is clearly governed by $m_D$. Note the presence of an imaginary part on the left. Solving with appropriate boundary conditions $\left.{\rm Re}V_C(r)\right|_{r=\infty}=0$, $\left.{\rm Im}V_C(r)\right|_{r=0}=0$ as well as $\left.\partial_r {\rm Im}V_C(r)\right|_{r=\infty}=0$ consistently gives Eq.\eqref{Eq:VHTL}. The important point is that in our approach the in-medium modification of the confining part of the vacuum potential also contributes. Since its Gauss law cannot be diagonalized by a simple Fourier transform we argue instead in \cite{Burnier:2015nsa}, based on comparison with the outcome if a simple background charge density is assumed, that its defining equation will also have a linear response form with the same l.h.s. as in Eq.\eqref{Eq:Coulomb(r)withIm}. Its in-medium modification is then governed by the parameter $\mu^4=m_D^2\frac{\sigma}{\tilde\alpha_s}$ and leads to 
\beq
-\frac{1}{r^{2}}\frac{d^2 V_s(r)}{dr^2}+ \mu^4 V_s(r)=\sigma \Big( 4 \pi \delta(\vr)- i T m_D^2 \varphi(m_D r)\Big).\label{Eq:String(r)withIm}
\eeq
This expression can still be solved explicitly using the parabolic cylinder functions $D_\nu(x)$
\begin{align}
\hspace{-0.3cm} {\rm Re}[V_s](r)&=-\frac{\Gamma[\frac{1}{4}]}{2^{\frac{3}{4}}\sqrt{\pi}}\frac{\sigma}{\mu} D_{-\frac{1}{2}}\big(\sqrt{2}\mu r\big)+ \frac{\Gamma[\frac{1}{4}]}{2\Gamma[\frac{3}{4}]} \frac{\sigma}{\mu}, \;\; {\rm Im}[V_s](r)=-i\frac{\sigma m_D^2 T}{\mu^4} \psi(\mu r)=-i\tilde\alpha_s T \psi(\mu r)\label{Eq:ImVSGenGauss},
\end{align}
where the expression $\psi(x)$ amounts to a Wronskian construction
\begin{eqnarray} 
&&\psi(x)=D_{-1/2}(\sqrt{2}x)\int_0^x dy\, {\rm Re}D_{-1/2}(i\sqrt{2}y)y^2 \varphi(ym_D/\mu)\\\notag&&+{\rm Re}D_{-1/2}(i\sqrt{2}x)\int_x^\infty dy\, D_{-1/2}(\sqrt{2}y)y^2 \varphi(ym_D/\mu) -D_{-1/2}(0)\int_0^\infty dy\, D_{-1/2}(\sqrt{2}y)y^2 \varphi(ym_D/\mu).
\end{eqnarray}
Ultimately we combine the above results by a simple linear combination ${\rm Re}[V]={\rm Re}[V_C]+{\rm Re}[V_s]$ and ${\rm Im}[V]={\rm Im}[V_C]+{\rm Im}[V_s]$.

\begin{figure*}[t]
\includegraphics[scale=0.35, trim= 2.5cm 0 2.5cm 0, clip=true ]{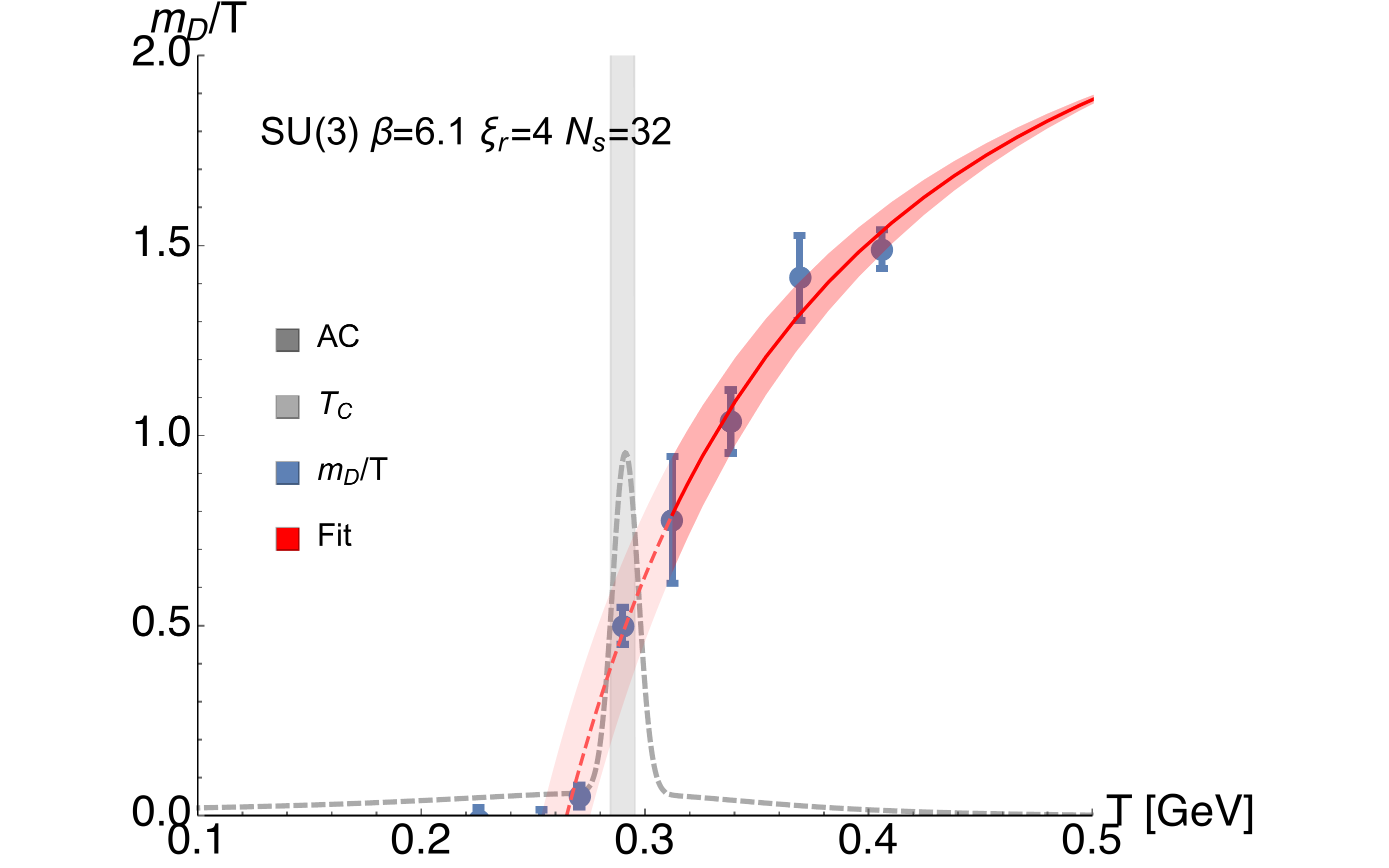}
\includegraphics[scale=0.35, trim= 2.8cm 0 2.5cm 0, clip=true ]{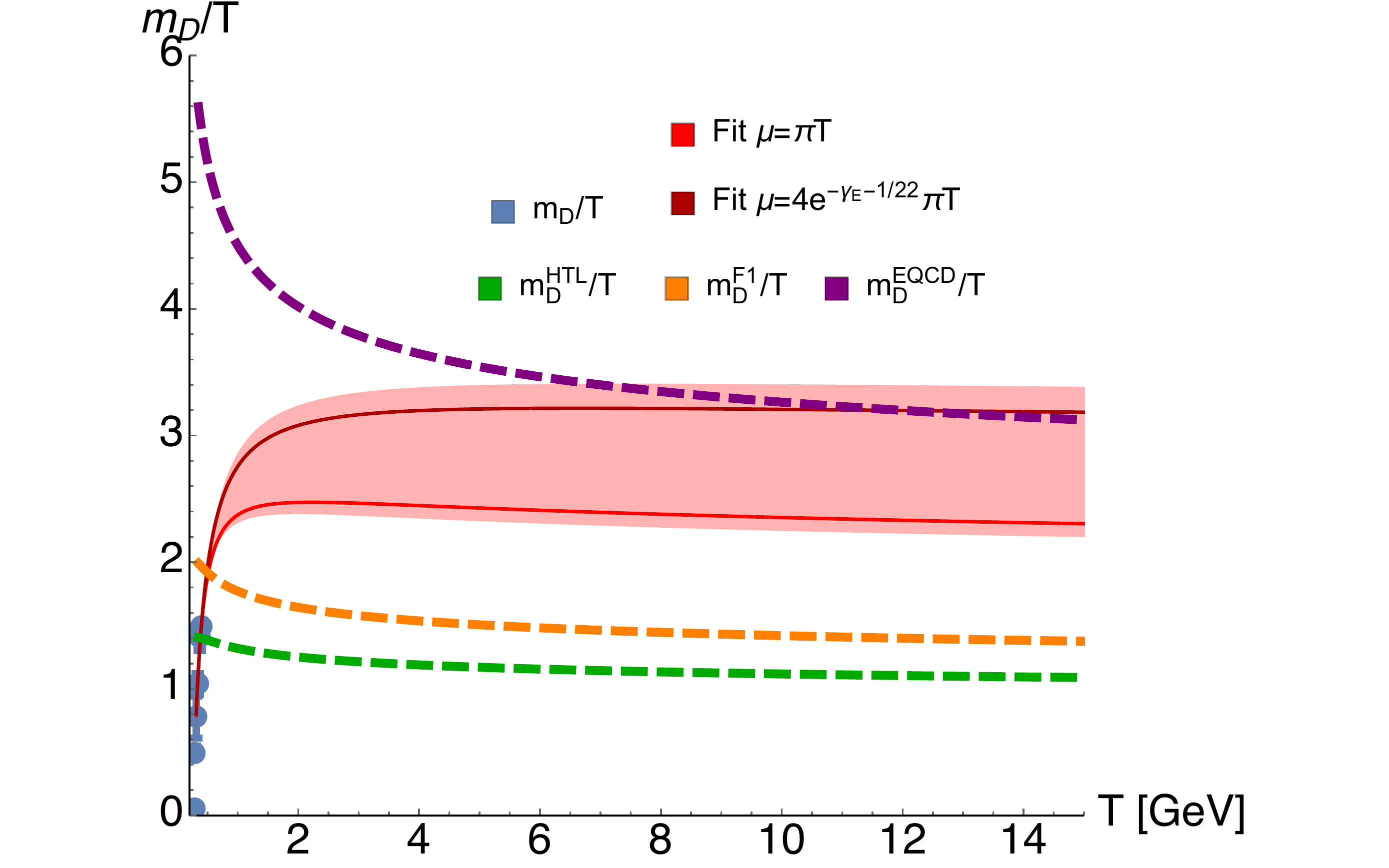}
\caption{(left) Debye mass parameter $m_D$ extracted from quenched QCD (blue points) together with a continuum corrected fit (red) based on the points above $T_C$. $m_D$ takes on finite values at $T_C$, identified by the gray band. As a crosscheck for the location of $T_C$, we also plot the normalized autocorrelation time of a generic Wilson line correlator, which peaks at this temperature. (right) Comparison of the extrapolation of our values of $m_D$ (red band) to values from the literature, i.e. using pure HTL (green), color singlet free energies (orange) and EQCD (purple). The bright red solid curve correspond to $\mu=\pi T$ the darker red one to $\mu=4e^{-\gamma_E-1/22}\pi T$. }\label{Fig:DebMass}
\end{figure*}

Let us see how well our ansatz allows us to reproduce the lattice results for the potential. As we cannot simulate directly at $T=0$, but do not find any changes in ${\rm Re}[V]$ below $T_C$, we fix the vacuum parameters of $V_{\rm T=0}$ using our data at $T=105$MeV. Note that once $\alpha_S$, $\sigma$ and $c$ are set via ${\rm Re}[V]$ at $T\approx0$ all in-medium modification of both ${\rm Re}[V]$ and ${\rm Im}[V]$ is characterized by the single T-dependent parameter $m_D$. We continue by fitting ${\rm Re}[V]$ via tuning of $m_D$, the outcome of which is shown as the solid  curves on the left of Fig.\ref{SU3Pot} and the corresponding values of $m_D$ are given as blue points in Fig.\ref{Fig:DebMass}. We find that our analytic parametrization of ${\rm Re}[V]$ works excellently and manages to faithfully retrace the lattice points within their uncertatinties.

Once $m_D$ is set, also ${\rm Im}[V]$ in our approach is fixed and provides a postdiction of the lattice values. As seen on the right of Fig.\ref{SU3Pot} at high temperatures and small distances there is even quantitative agreement with the lattice, while at lower temperatures we find  larger deviations. On the one hand it is clear that close to and below $T_C$ the ansatz of a perturbative medium we made in the derivation will become less and less justified and on the other hand it has to be kept in mind that the spectral width on which ${\rm Im}[V]$ is based was extracted using ${\cal O}(20)$ datapoints, which is quite small.

As shown on the left of Fig.\ref{Fig:DebMass}, the values of $m_D/T$ match the behavior found from inspecting ${\rm Re}[V]$. Where there is no modification of the real-part below $T_C$ the mass is compatible with zero and quickly rises to a finite value at $T_C$. Our expectation is that if the thermodynamic limit is taken one will observe a genuine jump due to the first order nature of the ensuing phase transition. The red solid line denotes a fit according to Eq.\eqref{Eq:mDpert} of the points above $T_C$ (i.e. within the dark red region). Here we assume that all higher order corrections in $g$ are captured in the two additional fit parameters $\kappa_{1,2}$. In the absence of a genuine continuum extrapolation the fit uses a continuum corrected $m_D^{\rm corr}=(t=T/T_C^{\rm phys})=\frac{m_D\left(t\right)}{\sqrt{\sigma_{\rm latt}}}\sqrt{\sigma^{\rm cont}}$ with $T_C^{\rm phys}=0.271$GeV and $\sigma_{\rm phys}=0.173$GeV. The four-loop running coupling we deploy is that of \cite{Vermaseren:1997fq} and evaluated at the scale $\mu=\pi T$. Coincidentally the fit manages to capture even the lowest non-vanishing data.The values for the non-perturbative corrections to $m_D$ obtained are a relatively large $\kappa=2.53\pm0.25$ and $\kappa_2=-1.41\pm0.15$, the latter is in particular necessary to reproduce the downward trend towards zero close to $T=T_C$.

Compared to previous results in the literature, we find distinct differences. Both the pure HTL expression of $m_D/T$ as well as the values extracted from the color singlet free energies show an upward bending close to $T_C$, while we observe a clear trend towards zero. Furthermore the slope of our extracted values indicates that they will rise above those of the previously mentioned results at intermediate $T>T_C$. If we extrapolate the fit naively, indeed we would find values almost twice those of pure HTL at $T\sim 10$GeV. Of course a large uncertainty lies within this extrapolation, which we may make more explicit by fitting our $m_D/T$ using the different scale setting $\mu=4e^{-\gamma_E-1/22}\pi T$, with $\gamma_E$ the Euler constant, which was used in the study of $m_D$ from EQCD. This fit gives less natural values for $\kappa_{1,2}$ but equally well reproduces our points around $T_C$. Its extrapolation however leads to a behavior at $T\sim10$GeV that becomes compatible with that found in EQCD itself. 

It would be very interesting to reduce this uncertainty by extracting the in-medium potential at much higher temperatures than those considered here. One in turn would be able to ascertain which scenario is eventually realized and whether $m_D$ as defined here will stay far from the perturbative predictions similar as is the case in EQCD. At the same time a continuum extrapolation of the underlying complex valued heavy-quark potential at finite temperature will eventually be required to connect unambiguously to the values of $m_D$ in perturbation theory at high temperature. A first extraction of $m_D$ in the context of studying quarkonium binding has been performed also in full QCD based on $N_f=2+1$ asqtad lattices \cite{Burnier:2015tda,Burnier:2016kqm} and its determination on more realistic HISQ ensembles in work in progress. On the conceptual side it needs to be better understood whether the $m_D$ considered here can be more formally connected to the concept of a thermal gluon mass, which would enable its use beyond the description of heavy-quarkonium binding properties at finite temperature.

Calculations for \cite{Burnier:2016mxc} were performed on the in-house cluster at the ITP in Heidelberg, the SuperB cluster at EPFL and the BWUniCluster at the KIT. YB is supported by SNF grant PZ00P2-142524. This work is part of and supported by the DFG Collaborative Research Centre "SFB 1225 (ISOQUANT)".


\begin{thebibliography}{99}
  
%\cite{Arnold:1995bh}
\bibitem{Arnold:1995bh} 
  P.~B.~Arnold and L.~G.~Yaffe,
  %``The NonAbelian Debye screening length beyond leading order,''
  Phys.\ Rev.\ D {\bf 52}, 7208 (1995)
  %[hep-ph/9508280].
  %%CITATION = HEP-PH/9508280;%%
  %143 citations counted in INSPIRE as of 24 juin 2015   
  
%\cite{Kajantie:1997pd,Hart:2000ha,Laine:2009dh}
%\cite{Kajantie:1997pd}
\bibitem{Kajantie:1997pd} 
  K.~Kajantie, M.~Laine, J.~Peisa, A.~Rajantie, K.~Rummukainen and M.~E.~Shaposhnikov,
  %``Nonperturbative Debye mass in finite temperature QCD,''
  Phys.\ Rev.\ Lett.\  {\bf 79}, 3130 (1997)
 % [hep-ph/9708207].
  %%CITATION = HEP-PH/9708207;%%
  %113 citations counted in INSPIRE as of 25 Jun 2015
\bibitem{Hart:2000ha} 
  A.~Hart, M.~Laine and O.~Philipsen,
  %``Static correlation lengths in QCD at high temperatures and finite densities,''
  Nucl.\ Phys.\ B {\bf 586}, 443 (2000)
 % [hep-ph/0004060].
  %%CITATION = HEP-PH/0004060;%%
  %143 citations counted in INSPIRE as of 25 juin 2015
%\cite{Laine:2009dh}
\bibitem{Laine:2009dh} 
  M.~Laine and M.~Vepsalainen,
  %``On the smallest screening masses in hot QCD,''
  JHEP {\bf 0909}, 023 (2009)
 % [arXiv:0906.4450 [hep-ph]].
  %%CITATION = ARXIV:0906.4450;%%
  %15 citations counted in INSPIRE as of 25 juin 2015

\bibitem{DebyHueckel}
P.~Debye, H.~H\"uckel,
Physikalische Zeitschrift 24, 185-206 (1923).

%\cite{Maezawa:2007fc}\cite{Digal:2005ht}
\bibitem{Maezawa:2007fc} 
  Y.~Maezawa {\it et al.}  [WHOT-QCD Collaboration],
  %``Heavy-quark free energy, debye mass, and spatial string tension at finite temperature in two flavor lattice QCD with Wilson quark action,''
  Phys.\ Rev.\ D {\bf 75}, 074501 (2007),
  Prog.\ Theor.\ Phys.\  {\bf 128}, 955 (2012).
 % [arXiv:1112.2756 [hep-lat]].
  %%CITATION = ARXIV:1112.2756;%%
  %12 citations counted in INSPIRE as of 25 juin 2015  
%\cite{Digal:2005ht}
\bibitem{Digal:2005ht} 
  S.~Digal, O.~Kaczmarek, F.~Karsch and H.~Satz,
  %``Heavy quark interactions in finite temperature QCD,''
  Eur.\ Phys.\ J.\ C {\bf 43}, 71 (2005)
  %[hep-ph/0505193].
  %%CITATION = HEP-PH/0505193;%%
  %38 citations counted in INSPIRE as of 24 juin 2015  
  
%\cite{Burnier:2016mxc}
\bibitem{Burnier:2016mxc} 
  Y.~Burnier and A.~Rothkopf,
  %``Complex heavy-quark potential and Debye mass in a gluonic medium from lattice QCD,''
  arXiv:1607.04049 [hep-lat].
  %%CITATION = ARXIV:1607.04049;%%
  %1 citations counted in INSPIRE as of 23 Oct 2016

%\cite{Burnier:2015nsa}
\bibitem{Burnier:2015nsa} 
  Y.~Burnier and A.~Rothkopf,
  %``A gauge invariant Debye mass and the complex heavy-quark potential,''
  Phys.\ Lett.\ B {\bf 753}, 232 (2016)
  %doi:10.1016/j.physletb.2015.12.031
  %[arXiv:1506.08684 [hep-ph]].
  %%CITATION = doi:10.1016/j.physletb.2015.12.031;%%
  %9 citations counted in INSPIRE as of 23 Oct 2016

%\cite{Brambilla:1999xf}\cite{Brambilla:2011sg}
\bibitem{Brambilla:1999xf}
  N.~Brambilla, A.~Pineda, J.~Soto and A.~Vairo,
  %``Potential NRQCD: An Effective theory for heavy quarkonium,''
  Nucl.\ Phys.\ B {\bf 566} (2000) 275,
  %[hep-ph/9907240].
  %%CITATION = HEP-PH/9907240;%%
  %406 citations counted in INSPIRE as of 08 Oct 2015
%\cite{Brambilla:2000gk}\cite{Pineda:2000sz}
%\bibitem{Brambilla:2000gk}
  %N.~Brambilla, A.~Pineda, J.~Soto and A.~Vairo,
  %``The QCD potential at O(1/m),''
  Phys.\ Rev.\ D {\bf 63} (2001) 014023.
  %[hep-ph/0002250].
  %%CITATION = HEP-PH/0002250;%%
  %121 citations counted in INSPIRE as of 08 Oct 2015
%\cite{Pineda:2000sz}
\bibitem{Pineda:2000sz}
  A.~Pineda and A.~Vairo,
  %``The QCD potential at O (1 / $m^{2)}$ : Complete spin dependent and spin independent result,''
  Phys.\ Rev.\ D {\bf 63} (2001) 054007
   %[Phys.\ Rev.\ D {\bf 64} (2001) 039902]
 % [hep-ph/0009145].
  %%CITATION = HEP-PH/0009145;%%
  %109 citations counted in INSPIRE as of 08 Oct 2015
%\cite{Brambilla:2011sg}
\bibitem{Brambilla:2011sg}
  N.~Brambilla, M.~A.~Escobedo, J.~Ghiglieri and A.~Vairo,
  %``Thermal width and gluo-dissociation of quarkonium in pNRQCD,''
  JHEP {\bf 1112} (2011) 116,
  %[arXiv:1109.5826 [hep-ph]].
  %%CITATION = ARXIV:1109.5826;%%
  %33 citations counted in INSPIRE as of 09 Oct 2015  
%\cite{Brambilla:2013dpa}
%\bibitem{Brambilla:2013dpa}
  %N.~Brambilla, M.~A.~Escobedo, J.~Ghiglieri and A.~Vairo,
  %``Thermal width and quarkonium dissociation by inelastic parton scattering,''
  JHEP {\bf 1305} (2013) 130.
  %[arXiv:1303.6097 [hep-ph]].
  %%CITATION = ARXIV:1303.6097;%%
  %25 citations counted in INSPIRE as of 09 Oct 2015

%\cite{Laine:2006ns}
\bibitem{Laine:2006ns}
  M.~Laine, O.~Philipsen, P.~Romatschke and M.~Tassler,
  %``Real-time static potential in hot QCD,''
  JHEP {\bf 0703}, 054 (2007)
  %[hep-ph/0611300].
  %%CITATION = HEP-PH/0611300;%%
  %208 citations counted in INSPIRE as of 15 sept. 2015 
\bibitem{Beraudo:2007ky} 
  A.~Beraudo, J.~-P.~Blaizot, C.~Ratti,
  %``Real and imaginary-time Q anti-Q correlators in a thermal medium,''
  Nucl.\ Phys.\ A {\bf 806}, 312 (2008) % [arXiv:0712.4394 [nucl-th]].
  %%CITATION = ARXIV:0712.4394;%%
  %93 citations counted in INSPIRE as of 29 Jun 2013  


 %\cite{Rothkopf:2011db}
\bibitem{Rothkopf:2009db} 
  A.~Rothkopf, T.~Hatsuda and S.~Sasaki,
    %``Proper heavy-quark potential from a spectral decomposition of the thermal Wilson loop,''
  PoS LAT {\bf 2009}, 162 (2009).
\bibitem{Rothkopf:2011db} 
  A.~Rothkopf, T.~Hatsuda and S.~Sasaki,
  %``Complex Heavy-Quark Potential at Finite Temperature from Lattice QCD,''
  Phys.\ Rev.\ Lett.\  {\bf 108}, 162001 (2012)
  %[arXiv:1108.1579 [hep-lat]].
  %%CITATION = ARXIV:1108.1579;%%
  %70 citations counted in INSPIRE as of 24 juin 2015
  
\bibitem{Burnier:2012az}
  Y.~Burnier and A.~Rothkopf,
  %``Disentangling the timescales behind the non-perturbative heavy quark potential,''
  Phys.\ Rev.\ D {\bf 86} (2012) 051503,
    %``A hard thermal loop benchmark for the extraction of the nonperturbative $Q\bar{Q}$ potential,''
  Phys.\ Rev.\ D {\bf 87}, 114019 (2013).
 %[arXiv:1208.1899 [hep-ph]].
  %%CITATION = ARXIV:1208.1899;%%
  %22 citations counted in INSPIRE as of 14 Aug 2014  
  
\bibitem{Burnier:2013nla}
  Y.~Burnier and A.~Rothkopf,
  %``Bayesian Approach to Spectral Function Reconstruction for Euclidean Quantum Field Theories,''
  Phys.\ Rev.\ Lett.\  {\bf 111} (2013) 18,  182003 % [arXiv:1307.6106 [hep-lat]].
  %%CITATION = ARXIV:1307.6106;%%
  %14 citations counted in INSPIRE as of 07 May 2014  
 
%\cite{Burnier:2014ssa}
\bibitem{Burnier:2014ssa} 
  Y.~Burnier, O.~Kaczmarek and A.~Rothkopf,
  %``Static quark-antiquark potential in the quark-gluon plasma from lattice QCD,''
  Phys.\ Rev.\ Lett.\  {\bf 114}, no. 8, 082001 (2015)
  %[arXiv:1410.2546 [hep-lat]].
  %%CITATION = ARXIV:1410.2546;%%
  %13 citations counted in INSPIRE as of 24 Jun 2015  
 
 %\cite{Matsufuru:2001cp}
  \bibitem{Matsufuru:2001cp}
  H.~Matsufuru {\it et al.},
  %``Numerical study of O(a) improved Wilson quark action on anisotropic lattice,''
  Phys.\ Rev.\  {\bf D64}, 114503 (2001).
  %[hep-lat/0107001]. 
 
%\cite{Dixit:1989vq}
\bibitem{Dixit:1989vq} 
  V.~V.~Dixit,
  %``Charge Screening and Space Dimension,''
  Mod.\ Phys.\ Lett.\ A {\bf 5}, 227 (1990).
  %%CITATION = MPLAE,A5,227;%%
  %31 citations counted in INSPIRE as of 24 juin 2015    
  
%\cite{Thakur:2013nia}
\bibitem{Thakur:2013nia} 
  L.~Thakur, U.~Kakade and B.~K.~Patra,
  %``Dissociation of Quarkonium in a Complex Potential,''
  Phys.\ Rev.\ D {\bf 89}, no. 9, 094020 (2014)
  %[arXiv:1401.0172 [hep-ph]].
  %%CITATION = ARXIV:1401.0172;%%
  %8 citations counted in INSPIRE as of 24 juin 2015    

%\cite{Vermaseren:1997fq}
\bibitem{Vermaseren:1997fq} 
  J.~A.~M.~Vermaseren, S.~A.~Larin and T.~van Ritbergen,
  %``The four loop quark mass anomalous dimension and the invariant quark mass,''
  Phys.\ Lett.\ B {\bf 405}, 327 (1997)
  %[hep-ph/9703284].
  %%CITATION = HEP-PH/9703284;%%
  %374 citations counted in INSPIRE as of 21 sept. 2015

%\cite{Burnier:2015tda}\cite{Burnier:2016kqm}
\bibitem{Burnier:2015tda} 
  Y.~Burnier, O.~Kaczmarek and A.~Rothkopf,
  %``Quarkonium at finite temperature: Towards realistic phenomenology from first principles,''
  JHEP {\bf 1512}, 101 (2015)
  %doi:10.1007/JHEP12(2015)101
  %[arXiv:1509.07366 [hep-ph]].
  %%CITATION = doi:10.1007/JHEP12(2015)101;%%
  %11 citations counted in INSPIRE as of 23 Oct 2016
%\cite{Burnier:2016kqm}
\bibitem{Burnier:2016kqm} 
  Y.~Burnier, O.~Kaczmarek and A.~Rothkopf,
  %``In-medium P-wave quarkonium from the complex lattice QCD potential,''
  JHEP {\bf 1610}, 032 (2016)
  %doi:10.1007/JHEP10(2016)032
  %[arXiv:1606.06211 [hep-ph]].
  %%CITATION = doi:10.1007/JHEP10(2016)032;%%
  
\end{thebibliography}
\end{document}